\documentclass[
reprint, 
longbibliography,
superscriptaddress,
preprintnumbers,
nobibnotes,
amsmath,
amssymb,
aps,
pra,
floatfix
]{revtex4-2}

\usepackage{graphicx}
\usepackage{dcolumn}
\usepackage{bm}
\usepackage{floatrow}
\usepackage[%
  colorlinks=true,
  urlcolor=blue,
  linkcolor=blue,
  citecolor=blue
]{hyperref}

\hypersetup{breaklinks=true} 

\usepackage{float}
\usepackage{dcolumn}
\usepackage{bm}
\usepackage{xcolor}
\usepackage{braket}
\usepackage{upgreek} 

\usepackage{tabularray}

\usepackage{siunitx} 

\usepackage{comment}
\usepackage{graphicx} 
\begin{document}


\title{Ramsey interferometry of nuclear spins in diamond\\ using stimulated Raman adiabatic passage}

\author{Sean Lourette}\email{slourette@berkeley.edu}
\affiliation{Department of Physics, University of California, Berkeley, California 94720, USA}
\affiliation{DEVCOM Army Research Laboratory, Adelphi, Maryland 20783, USA}  

\author{Andrey Jarmola}
\affiliation{Department of Physics, University of California, Berkeley, California 94720, USA}
\affiliation{DEVCOM Army Research Laboratory, Adelphi, Maryland 20783, USA} 

\author{Jabir Chathanathil}
\affiliation{DEVCOM Army Research Laboratory, Adelphi, Maryland 20783, USA} 

\author{Sebasti\'an C. Carrasco}
\affiliation{DEVCOM Army Research Laboratory, Adelphi, Maryland 20783, USA}

\author{Dmitry Budker}
\affiliation{Department of Physics, University of California, Berkeley, California 94720, USA}
\affiliation{Helmholtz-Institut Mainz, 55099 Mainz, Germany}
\affiliation{GSI Helmholtzzentrum für Schwerionenforschung GmbH, 64291 Darmstadt, Germany}
\affiliation{Johannes Gutenberg-Universit{\"a}t Mainz, 55128 Mainz, Germany}

\author{Svetlana A. Malinovskaya}
\affiliation{Helmholtz-Institut Mainz, 55099 Mainz, Germany}
\affiliation{GSI Helmholtzzentrum für Schwerionenforschung GmbH, 64291 Darmstadt, Germany}
\affiliation{Johannes Gutenberg-Universit{\"a}t Mainz, 55128 Mainz, Germany}
\affiliation{Department of Physics, Stevens Institute of Technology, Hoboken, NJ 07030}
\affiliation{Technische Universit{\"a}t Darmstadt, Institut f{\"u}r Angewandte Physik, D-64289 Darmstadt, Germany}

\author{A. Glen Birdwell} 
\affiliation{DEVCOM Army Research Laboratory, Adelphi, Maryland 20783, USA} 

\author{Tony G. Ivanov}
\affiliation{DEVCOM Army Research Laboratory, Adelphi, Maryland 20783, USA} 

\author{Vladimir S. Malinovsky}
\affiliation{DEVCOM Army Research Laboratory, Adelphi, Maryland 20783, USA}

\date{\today}

\begin{abstract}

We report the first experimental demonstration of stimulated Raman adiabatic passage (STIRAP) in nuclear-spin transitions of $^{14}$N within nitrogen-vacancy (NV) color centers in diamond.
It is shown that the STIRAP technique suppresses the occupation of the intermediate state, which is a crucial factor for improvements in quantum sensing technology.
Building on that advantage, we develop and implement a generalized version of the Ramsey interferometric scheme, employing half-STIRAP pulses to perform the necessary quantum-state manipulation with high fidelity.
The enhanced robustness of the STIRAP-based Ramsey scheme to variations in the pulse parameters is experimentally demonstrated, showing good agreement with theoretical predictions.
Our results pave the way for improving the long-term stability of diamond-based sensors, such as gyroscopes and frequency standards.
\end{abstract}

\maketitle

\section{Introduction} 

Quantum sensing signifies a groundbreaking development in detection methodologies. 
Leveraging the principles of quantum mechanics, it enables the creation of remarkably precise and sensitive detectors~\cite{YE2024}.
A typical quantum sensor involves the use of quantum particles (such as photons, atoms, or ions), the manipulation of these particles' states, and the precise measurement of changes in those states.
%
The highest precision atomic clocks employ a basic two-level (TL) atomic system as their core element,
with their superior performance being directly linked to the ability to accurately control atomic states~\cite{LUD2015}.


The TL atomic system driven by an electromagnetic field has proven to be surprisingly rich in physical phenomena~\cite{Allen_Eberly_book}. 
The quantum control of the TL wave function has been extensively studied and successfully implemented in various technological applications. The primary mechanism for controlling a TL system relies on the pulse area theorem, 
which is evidenced by Rabi oscillations of the state population as a function of the area of the applied pulses ~\cite{Allen_Eberly_book,BWShore,berman2011principles,Sola_AAMO2018}. 

However, it is often necessary to control more than just two states of a primary sensor particle. Even adding a third level to the system introduces multiple additional effects that cannot be fathomed in a simple TL particle.
Examples include electromagnetically induced transparency (EIT)~\cite{Kocharovskaya_Khanin,Harris1997_EIT_Review} and stimulated Raman adiabatic passage (STIRAP)~\cite{BERG1998,Bergmann_STIRAP_Review,JunYeSTIRAP,Bohm2021_NV_STIRAP,Gong2024_NV_STIRAP,CUB2005,SOL2022}. The latter allows robust transfer of populations among the three states even when one of the states is ``lossy''. STIRAP has numerous applications across various fields of science~\cite{STIRAP_Roadmap}.
The basic STIRAP scheme has been extended to multi-level systems~\cite{MalPRA97}, and to ``fractional STIRAP''~\cite{Vitanov1999_FSTIRAP}, which allows the control of coherence, and more recently to ``chirped STIRAP''~\cite{Band1994_CHIRAP,Chathanathil2023_Chirped_FSTIRAP}, which, in addition to controlling coherence, achieves state manipulation with high spectral resolution. Another extension is ``two-way STIRAP"~\cite{Genov2023_TwoWay}, which enables robust \textit{swap} of populations between two states. Quantum control of entanglement applying STIRAP and fractional STIRAP has been considered in~\cite{MalPRA2004a,MalPRA2004b} showing robustness of the adiabatic method for entangled states generation. 

In this work, we utilize  pulse-area and STIRAP control methods to manipulate the quantum states of the $^{14}$N nuclear spins ($I=1$) within  nitrogen-vacancy (NV) color centers in diamond. 
Since  the $^{14}$N nuclear spin states form a three-level $\Lambda$ system, it is natural to consider STIRAP pulse sequences to control spin dynamics.
We develop an adiabatic extension of advanced Ramsey interferometry based on the STIRAP protocol.
We compare the performance of conventional and STIRAP-based Ramsey and demonstrate the advantages of the latter, particularly its robustness against variations in pulse parameters.
The presented results offer practical advantages and will facilitate further advancements in quantum sensing science and technology. 

This study, along with previous work by others~\cite{JAIN2004240}, demonstrates the successful application of STIRAP techniques in the field of nuclear magnetic resonance (NMR). Notably, this work reverses the typical trend, as NMR techniques are often adapted for use in atomic and molecular physics~\cite{Hahn1996}. Here, we showcase the successful transfer of a technique from atomic and molecular physics to the NMR domain, highlighting its  versatility and potential for cross-disciplinary innovation.


\begin{figure*}
    \centering
    \includegraphics[width=1.0\textwidth]{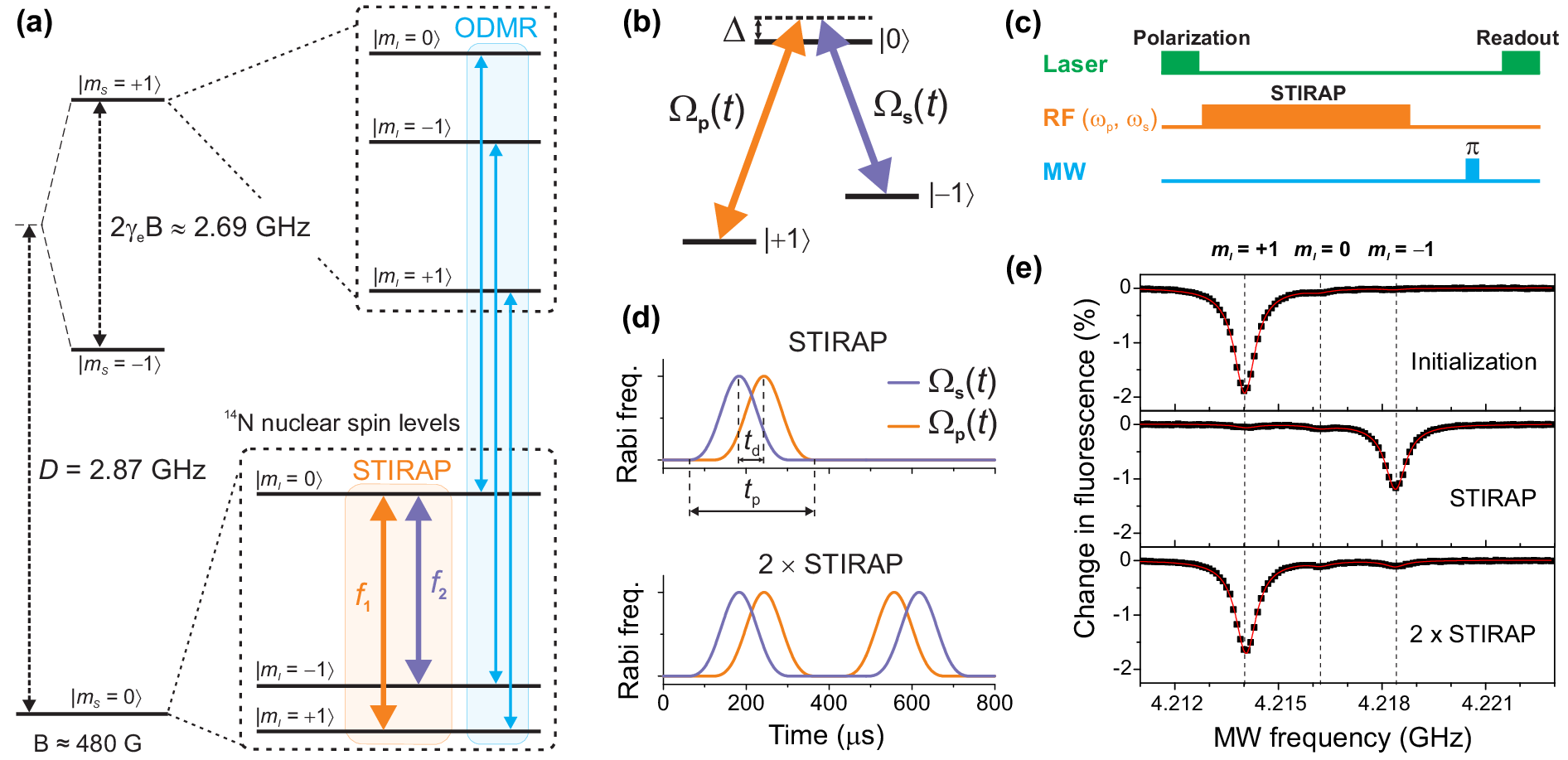}
    \caption{\label{fig:EnergyLevels} \textbf{STIRAP in $^{14}\mathrm{N}$ nuclear spins intrinsic to NV centers}
    \textbf{(a)}~NV ground-state energy levels. STIRAP is performed on the $^{14}\mathrm{N}$ nuclear spin triplet in the $m_S = 0$ electron spin state using RF pulses with frequencies near the transition frequencies $f_1$ $(\SI{5.089}{MHz})$ and $f_2$ $(\SI{4.797}{MHz})$. The readout of the nuclear spin is performed either through  ODMR using microwaves on the $\ket{m_S=0}\Leftrightarrow \ket{m_S=+1}$ transitions or through direct optical readout.
    \textbf{(b)}~Energy level diagram of the $^{14}\mathrm{N}$ nuclear-spin sublevels within $\ket{m_S=0}$ depicting the Rabi frequencies for the Stokes $\big(\Omega_s(t)\big)$ and pump $\big(\Omega_p(t)\big)$ pulses and the one-photon detuning~$\Delta$.
    \textbf{(c)}~Pulse sequence diagram showing the readout of the nuclear spin state using ODMR.
    \textbf{(d)}~Pulse shapes for pump and Stokes pulses for STIRAP and $2\times\mathrm{STIRAP}$.
    \textbf{(e)}~Fidelity of STIRAP, demonstrated for $\Delta = 0$. Pulsed ODMR is used to measure the populations of the nuclear-spin sublevels after initialization, STIRAP, and $2\times\mathrm{STIRAP}$.
    Markers -- experimental measurements, solid lines -- Lorentzian fit, vertical dashed lines -- positions of the three hyperfine components.
    }
\end{figure*}

\section{\label{STIRAP}STIRAP in nuclear spins in diamond}

To demonstrate the STIRAP population transfer between $^{14}$N nuclear spin states, we use a custom-built epifluorescence microscopy setup (described in~\cite{Jarmola2021}) to measure optically detected magnetic resonances (ODMRs) in an ensemble of NV centers. 
The diamond used for our experiments is chemical-vapor-deposition (CVD) grown, $^{12}\mathrm{C}$ enriched (99.99\% $^{12}\mathrm{C}$) [110] cut diamond plate, with initial nitrogen concentration of $\sim \SI{13}{ppm}$ and NV concentration of  $\sim\SI{4}{ppm}$.
A bias magnetic field $B$ (\SI{480}{G}) was applied along one of the NV axes using Halbach array composed of temperature-compensated samarium-cobalt (SmCo) magnets, designed to minimize the magnetic field gradients across the volume
($\sim \qtyproduct{50 x 50 x 150}{\micro m}$)
of interrogated NV centers (see Appendix~\ref{sec:Appendix:Experiment}).

Figure~\ref{fig:EnergyLevels}(a) shows the electron NV ground state triplet energy levels ($m_S = 0,\pm1$) and the $^{14}\mathrm{N}$ hyperfine sublevels ($m_I = 0,\pm1$) of $\ket{m_S=0}$ and $\ket{m_S=+1}$ for a magnetic field $B = \SI{480}{G}$ aligned to the NV axis.
At this magnetic field, the electron and nuclear spins are optically polarized into the $\ket{m_S,m_I} = \ket{0,+1}$ state, as a result of the excited-state level anticrossing (ESLAC).
This phenomenon is caused by an electro-nuclear spin-conserving flip-flop interaction, allowing the electron-spin polarization to be effectively transferred to the $^{14}\mathrm{N}$ nuclear spin, and is explained in~\cite{JAC2009,SME2009,STE2010PRB,FIS2013PRB}.

In this work, STIRAP (Fig.\,\ref{fig:EnergyLevels}(a), orange box) is performed on the $^{14}\mathrm{N}$ nuclear spins that are intrinsic to the NV center. Population is transferred from $m_I = +1$ to $m_I = -1$ in the $m_S = 0$ manifold using two radio-frequency (RF) fields with frequencies $\omega_p$ (of about $f_1 =$ $\SI{5.089}{MHz}$) and $\omega_s$ (of about $f_2 = $ $\SI{4.797}{MHz}$), and amplitudes $\Omega_p(t)$ and $\Omega_s(t)$, respectively.

The population of the three nuclear-spin sublevels is determined using optically detected magnetic resonance (ODMR) techniques~\cite{Bar2020,Dreau2011}.
There are three microwave (MW) transitions (Fig.\,\ref{fig:EnergyLevels}(a), cyan box) between $\ket{m_S=0}$ and $\ket{m_S=+1}$ that follow $\Delta m_I = 0$ selection rules, each transition corresponding to a particular nuclear spin state ($m_I$). The three transitions result in the appearance of three resonances ($\sim\SI{4.2}{GHz}$) in the ODMR spectrum that are used to determine the population of each $m_I$ state.
The relative populations of the nuclear spin state can also be measured using direct optical readout, which is feasible near ESLAC for reasons similar to those responsible for optical polarization~\cite{Jarmola2020Robust}.

\subsection{\label{sec:Theory} Sketch of STIRAP theory}

The Hamiltonian of the $^{14}\mathrm{N}$ nuclear spin states in the field-interaction representation under the rotating wave approximation (RWA)
in the case of two-photon resonance
has the form
\begin{align}\label{eq:Hamiltonian}
H (t)= \frac{\hbar}{2} 
\left(\begin{array}{ccc}
 0 & \Omega_p(t)  & 0 \\
\Omega_p(t)  & -2\Delta & \Omega_s(t) \\
0 & \Omega_s(t) & 0 
\end{array}\right) \,,
\end{align}
where $\Delta=\omega_p-\omega_1=\omega_s-\omega_2$ is the one-photon detuning shown in Fig.\,\ref{fig:EnergyLevels}(b), $\omega_{p,s}$ are the  frequencies of the pump and Stokes RF-fields, $\omega_{1,2}= 2\pi f_{1,2}$ are the frequencies of $\ket{m_{I}=1} \leftrightarrow \ket{m_{I}=0}$ and $\ket{m_{I}=0} \leftrightarrow \ket{m_{I}=-1}$ transitions depicted in Fig.\,\ref{fig:EnergyLevels}(a), $\Omega_{p,s}(t)=\gamma_n B_{p,s}(t)$ are the Rabi frequency envelopes of the pump and Stokes pulses.
Here, $B_{p,s}(t)$ refers to the amplitudes of the oscillations of the transverse magnetic field (perpendicular to the NV quantization axis) produced by the pump and Stokes fields, and $\gamma_n / 2 \pi = \SI{307.59(3)}{Hz \per G}$~\cite{Lourette2023} is the gyromagnetic ratio of the $^{14}\mathrm{N}$ nuclear spin in diamond.

The commonly accepted mechanism of the STIRAP protocol by which population is transferred from the initial state $\ket{1}$ to the target state $\ket{-1}$ can be demonstrated in the adiabatic basis. Diagonalizing the Hamiltonian in Eq.\,(\ref{eq:Hamiltonian}) reveals the three eigenstates
\begin{equation}\label{eq:dark}
\resizebox{0.9\hsize}{!}{$
    \ket{\mathcal{D}} \! = \!
    \begin{pmatrix}
        \cos\theta              \\
        0                           \\
        -\sin\theta               
    \end{pmatrix}
    \!\!,\,
    \ket{\mathcal{B}_-} \! = \!
    \begin{pmatrix}
        \sin\theta \sin\xi    \\
        \cos\xi                  \\
        \cos\theta \sin\xi
    \end{pmatrix}
    \!\!,\,
    \ket{\mathcal{B}_+} \! = \!
    \begin{pmatrix}
        \sin\theta \cos\xi    \\
        -\sin\xi                 \\
        \cos\theta \cos\xi
    \end{pmatrix}$
    \!\!,
}
\end{equation}
where time dependence of all terms has been omitted to simplify the notation,
$\tan \theta (t) =\Omega_p(t)/\Omega_s(t)$
and
$\tan 2 \xi (t) = \Omega_e(t)/\Delta$
define the mixing angles $\theta (t)$ and $\xi (t)$,
and $\Omega_e(t) = \sqrt{\Omega_p^2(t) + \Omega_s^2(t)}$ is the effective Rabi frequency~\cite{Shore:17}.
We define pulse area as $\mathcal{A} = \int \Omega_e(t) \, \mathrm{d}t$. 

From the expression of the so-called dark state $\ket{\mathcal{D}(t)}$ 
we can see that the ``counterintuitive'' or SP pulse sequence (the Stokes pulse precedes the pump pulse and it is turned off before the pump pulse ends) provides a possibility to transfer population from the state $\ket{1}$ to the state $\ket{-1}$. Indeed, if  $\Omega_p(t)/\Omega_s(t) \rightarrow 0$ ($\theta (t) \rightarrow 0$) the eigenstate $\ket{\mathcal{D}(t)}$ in Eq.\,(\ref{eq:dark}) correlates with state $\ket{1}$,  while if $\Omega_s(t)/\Omega_p(t) \rightarrow 0$ ($\theta (t) \rightarrow  \pi/2$) the eigenstate correlates with state $\ket{-1}$. To satisfy the transfer adiabaticity (to guarantee the system dynamics take place only in the zero eigenstate $\ket{\mathcal{D}(t)}$) and minimize residual population of the intermediate state $\ket{0}$ the pulses should be sufficiently strong and have substantial overlap~\cite{BERG1998}.

\subsection{\label{sec:STIRAP}STIRAP demonstration}

Figure~\ref{fig:EnergyLevels}(c) illustrates the pulse sequence utilized for manipulating nuclear spin states and subsequently measuring the populations of these states.
First, the $^{14}\mathrm{N}$ nuclear spins are optically initialized (i.e. polarized) into $m_I = +1$ using a \SI{532}{nm} laser pulse, making use of the ESLAC at \SI{480}{G}~\cite{JAC2009,SME2009,STE2010PRB,FIS2013PRB}.
Next, the state of nuclear spin is manipulated (e.g. STIRAP, $2\times\mathrm{STIRAP}$) using RF pulses with frequencies $\omega_p$ and $\omega_s$.
Then the population of each nuclear spin state is measured using pulsed ODMR~\cite{Bar2020}, in which the change in fluorescence is measured with and without a mapping microwave $\pi$-pulse (rectangular pulse of duration \SI{2}{\micro s}) whose frequency is scanned through the $\ket{m_S=0} \Leftrightarrow \ket{m_S=+1}$ transition.

To implement STIRAP transfer, we use the pump and Stokes pulses of the Blackman shape, defining the time dependent Rabi frequencies as
$\Omega_{p,s}(t) = \Omega_{0} w_B(t \mp t_d/2)$, where 
\begin{align}
    w_{B}(t) =
        \sin^2 \left(\pi t/T\right) - 0.16 \sin^2 \left(2 \pi t/T\right) \,,
\end{align}
for $ 0 \leq t \leq T$, $\Omega_0$ is the pulse amplitude, $T$ is the individual pulse duration, and $t_d$ is the offset parameter controlling the time delay between pulses. 
For all measurements performed in this work, we use $t_d = 0.25 T = 0.2 t_p$, where $t_p$ is defined to be the composite pulse duration, to minimize the nonadiabatic coupling between the dark state $\ket{\mathcal{D}(t)}$ and the bright states $\ket{\mathcal{B}_{\pm}(t)}$ of the Hamiltonian in Eq.\,(\ref{eq:Hamiltonian}).
For this particular construction of STIRAP pulses, the pulse area can be expressed as $\mathcal{A} \approx 0.5093 \, \Omega_{peak} \, t_p$, where $\Omega_{peak} = \left|\Omega_e(t)\right|_{\mathrm{max}} \approx 1.094 \, \Omega_{0}$ is the peak effective Rabi frequency.

To demonstrate the fidelity of STIRAP, we measure the pulsed-ODMR spectra (Fig.\,\ref{fig:EnergyLevels}(e), top to bottom):
$\textbf{(i)}$ after initialization (without RF pulses), in which the $^{14}\mathrm{N}$ nuclear spins are optically polarized into $\ket{m_I = +1}$,
$\textbf{(ii)}$ after STIRAP transfer (Fig.\,\ref{fig:EnergyLevels}(d)-top), when the population is transferred from $\ket{m_I = +1}$ to $\ket{m_I = -1}$, and 
$\textbf{(iii)}$ after $2\times\mathrm{STIRAP}$ transfer  (Fig.\,\ref{fig:EnergyLevels}(d)-bottom), when the population is transferred to $\ket{m_I = -1}$ (the first STIRAP) and then back to $\ket{m_I = +1}$ (second, or reverse STIRAP).
Note that the ODMR signal after STIRAP appears to be weaker than the signal after initialization (compare the top and the middle frames in Fig.\,\ref{fig:EnergyLevels}(e)).
This, however, does not indicate the loss of the population, but rather it reflects the difference of the relative brightness of the nuclear states~\cite{Jarmola2020Robust}.
Indeed, after STIRAP is performed a second time the signal largely returns to its original size, 95\% of its initially polarized value.

\begin{figure}
\centering
    \includegraphics[width=1\columnwidth]{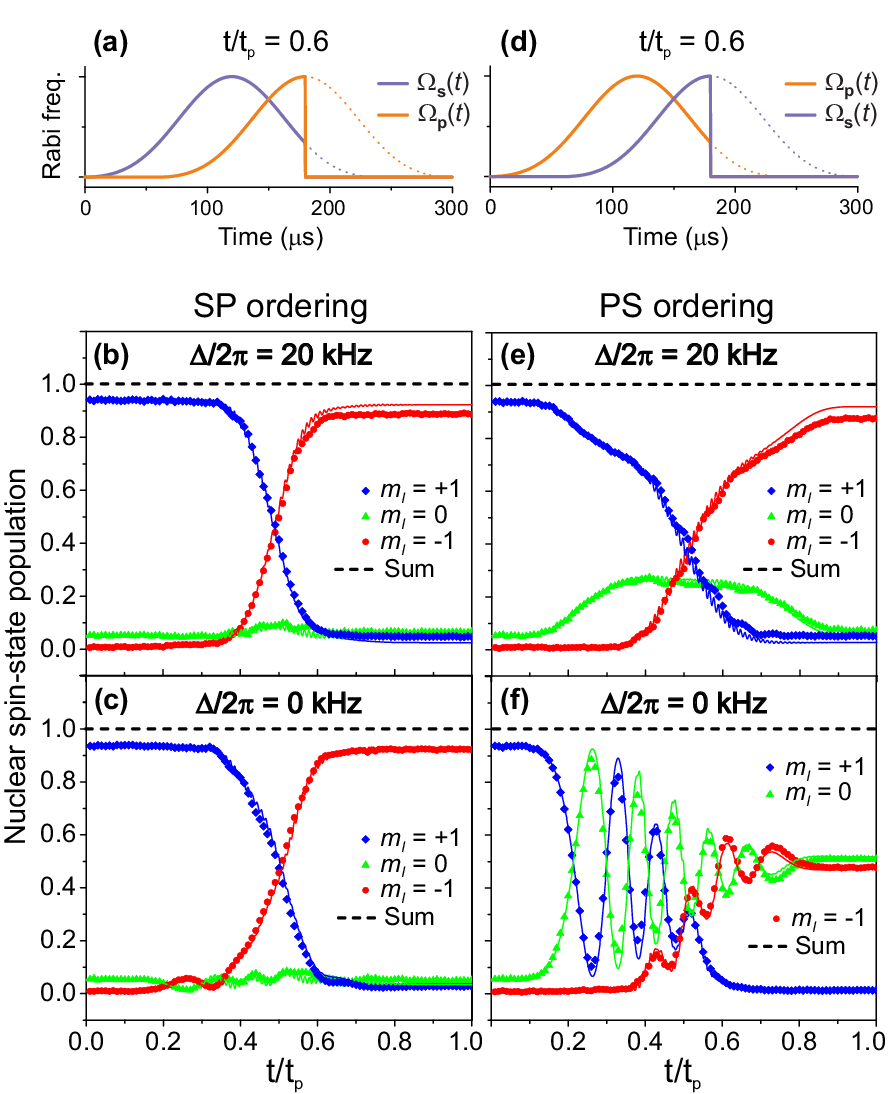}
    \caption{\label{fig:FSTIRAP} \textbf{STIRAP Dynamics.}
    \textbf{(a)}~Amplitudes of the applied truncated RF STIRAP Stokes and pump pulses, shown here for
    $t/t_p = 0.6,$
    which are used to determine the time-evolution of the $^{14}$N nuclear-spin-state population. The pulse area for the full non-truncated sequence is $\mathcal{A}/2 \approx 5.6 \pi$.
    \textbf{(b)},\textbf{(c)}~The time-evolution of the nuclear spin state during STIRAP for $\Delta/2\pi = \qtylist{0;20}{kHz}$, respectively.
    \textbf{(d)}~The same as \textbf{(a)}, but with the ordering of Stokes and pump pulse swapped (``intuitive'' or PS ordering).
    \textbf{(e)},\textbf{(f)}~The time-evolution of the nuclear spin state during the PS pulse sequence for $\Delta/2\pi = \qtylist{0;20}{kHz}$, respectively.
    Markers -- experimental measurements, solid lines -- theoretical model.
    }
\end{figure}

The dynamics of the $^{14}\mathrm{N}$ nuclear spin during STIRAP is measured by varying $t/t_p$, the fraction of the applied STIRAP pulse (Fig.\,\ref{fig:FSTIRAP}(a)), and subsequently measuring populations of $\ket{m_I}$ using pulsed ODMR.
In contrast with measurements shown in Fig.\,\ref{fig:EnergyLevels}(e), where the microwave frequency was scanned, here the microwave frequency was sequentially fixed to each of the three $\ket{m_S=0,m_I}\Leftrightarrow\ket{m_S=+1,m_I}$ transitions corresponding to the three hyperfine components.
This provides sufficient information to uniquely determine the nuclear spin state population.

\begin{figure*}
\centering
    \includegraphics[width=1.0\textwidth]{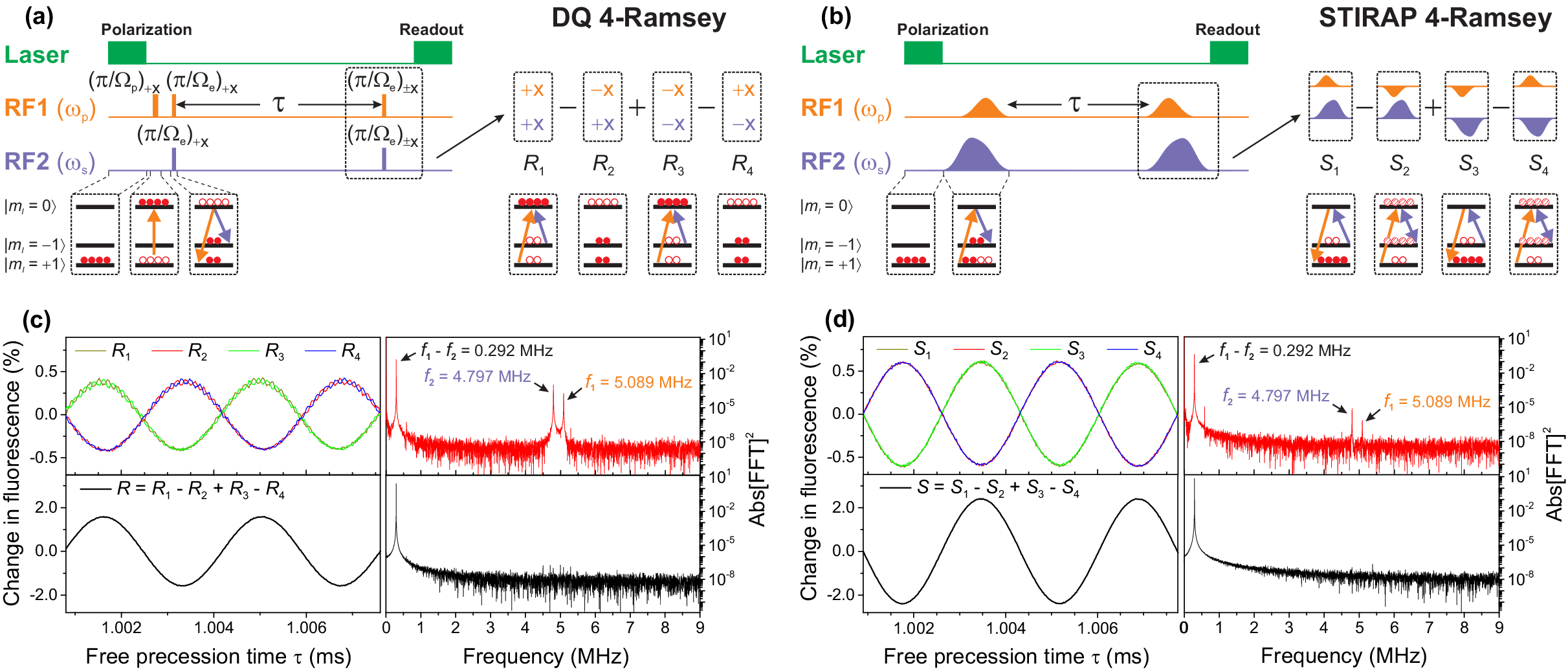}
    \caption{
        \label{fig:H_STIRAP}
        \textbf{Ramsey interferometry of $^{14}$N nuclear spins using STIRAP.}
        \textbf{(a)} The pulse sequence for nuclear double-quantum (DQ) 4-Ramsey.
        \textbf{(b)} The pulse sequence for nuclear STIRAP 4-Ramsey, using two half-STIRAP pulses.
        For both (a) and (b): the duration of the RF pulses are drawn to scale, and the final pulse is phase cycled according to the inset.
        \textbf{(c,d)} The experimentally measured Ramsey fringes obtained by scanning $\tau$ using the pulse sequences shown in (a) and (b) respectively.
        The four panels in (c) and (d) follow a similar structure:
        (Top-left) Ramsey fringes obtained by scanning $\tau$, for all four phase combinations
        and (Top-right) corresponding Fourier transforms.
        (Bottom-left) Linear combination of individual Ramsey measurements with different phases
        and (Bottom-right) corresponding Fourier transforms.
    }
\end{figure*}

Figure~\ref{fig:FSTIRAP}(b),(c) shows the experimentally measured (markers) and theoretically modeled (lines) time evolution of the nuclear spin state population during STIRAP transfer for resonant ($\Delta = 0$) and off-resonant ($\Delta/2\pi = 20$ kHz) conditions.
In both instances, the population is adiabatically transferred from
$\ket{m_I = +1}$ to $\ket{m_I = -1}$
through the dark state
$\ket{\mathcal{D}(t)}$ (Eq.\,(\ref{eq:dark})),
with a negligible transient population in the $\ket{m_I = 0}$ state.

The same measurements are repeated with the PS ordering of the pulses (pump then Stokes, as depicted in Fig.\,\ref{fig:FSTIRAP}(d)) resulting in nuclear spin state populations shown in Fig.\,\ref{fig:FSTIRAP}(e) and Fig.\,\ref{fig:FSTIRAP}(f) for $\Delta = 0$ and $\Delta/2\pi = 20$ kHz respectively.
For $\Delta/2\pi = \SI{20}{kHz}$, there is near-complete population transfer to $\ket{-1}$, while for $\Delta = 0$, the population is transferred to a nearly equal superposition of $\ket{-1}$ and $\ket{0}$.
These results can be understood in the adiabatic basis (in other words, by performing the dressed state analysis): in the case of the SP ordering, population is transferred exclusively through the dark state $\ket{\mathcal{D}(t)}$, but for the PS ordering the population is instead transferred through the two bright states $\ket{\mathcal{B}_{\pm}(t)}$.
For the PS ordering, the populations of these dressed states $\ket{\mathcal{B}_-(t)}$ and $\ket{\mathcal{B}_+(t)}$ at the initial time are defined by $\sin\xi (t=0)$ and $\cos\xi (t=0)$ respectively.
A deeper analysis shows that regardless of the value of $\Delta$, the population of the state $\ket{1}$ is always zero after the pulse is complete, assuming the adiabaticity condition is satisfied.
The details of the population dynamics at intermediate times depend on the effective pulse area $\mathcal{A}$ and the detuning value $\Delta$.

The dephasing rate in our sample is insufficient to produce the observed decay of the oscillations shown in Fig.\,\ref{fig:FSTIRAP}(f) on the timescale of the applied pulses.
In fact, this decay is explained by a gradient in RF amplitude, and therefore in
the effective Rabi frequency $\Omega_{e}(t)$,
across the interrogated volume of NV centers.
We measured the distribution of Rabi frequencies by performing Fourier analysis on Rabi oscillations of the $\ket{\pm 1} \leftrightarrow \ket{0}$ transitions (see Fig.\,\ref{fig:Distribution} in the Appendix).
We found the measured Rabi frequency distribution to be well described by a normal distribution.
All theoretical results have been obtained by averaging over the Rabi frequency distribution (see Appendix~\ref{sec:Appendix:Simulations}).

The modeling results in Fig.\,\ref{fig:FSTIRAP} are obtained using the pulse duration of $t_p = \SI{300}{\mu s}$ ($T=240 \mu$s, $t_d= 60 \mu$s), and the peak amplitude of the effective Rabi frequency $\Omega_e$ equal to  $\langle\Omega_{e} \rangle/2\pi = \SI{36.5}{kHz}$ and full width at half maximum of $0.164\,\left<\Omega_{e}\right>$, which agree with the experimentally measured values of $\SI{36.1}{kHz}$ and $0.166~\langle\Omega_{e} \rangle$.
We note the high frequency oscillations with a very small amplitude, visible in Fig.\,\ref{fig:FSTIRAP} on the modeling lines. 
These residual oscillations are a result of the ac Zeeman shifts from the pump field acting off-resonantly on $\ket{0} \leftrightarrow \ket{-1}$ and the Stokes field on $\ket{1} \leftrightarrow \ket{0}$ (see Appendix~\ref{sec:Appendix:Simulations} for the explanation).

\section{\label{sec:FSTIRAP} Ramsey interferometry using half-STIRAP}

Ramsey interferometry is a technique commonly used in sensing applications to precisely measure the transition frequencies of a quantum system.
In a basic Ramsey interferometry scheme, a pair of $\pi/2$ pulses delayed by a free evolution interval is applied to an ensemble of identical TL systems.
Ideally, the durations of the pulses are negligible, and the area of each pulse is exactly equal to $\pi/2$.
Under these conditions, the measured Ramsey fringes are not sensitive to the pulse parameters (such as detuning), allowing for the detection of an external perturbation to the transition frequency through the phase shift of the fringes. In reality, the pulses have finite durations and may have non-ideal areas. Additionally, a homogeneous excitation of the ensemble may pose an experimental challenge. As a result of the imperfections, the measured phase shifts are sensitive to the pulse parameters and must be taken into account during sensor development.

In rotation sensing with $^{14}\mathrm{N}$ nuclear spins that are intrinsic to NV centers~\cite{Jarmola2021,Soshenko2021}, the double-quantum (DQ) transition frequency
($\ket{m_S, m_I}$: $\ket{0,+1} \Leftrightarrow \ket{0,-1}$)
is measured using Ramsey interferometry, in which a pair of rectangular $\pi/2$ RF pulses are used (Fig.\,\ref{fig:H_STIRAP}a).
However, when using ensembles of NV centers, there is a distribution of Rabi frequencies arising from RF power gradients across the sensing volume, which limits the fidelity and robustness of the technique. 
To overcome these shortcomings, we develop a new Ramsey technique replacing the rectangular pulses with adiabatic pulses, (half-STIRAP, Fig.\,\ref{fig:H_STIRAP}b).
In the previous section, we demonstrate the STIRAP protocol showing the adiabatic manipulation of the $^{14}\mathrm{N}$ nuclear spin state population.
Here, we address the phase sensitivity of adiabatic control by creating an adiabatic version of the DQ Ramsey scheme and compare the performance of both techniques.

The STIRAP process demonstrated in the previous section can be modified to create a superposition of the initial $\ket{1}$ and the target state $\ket{-1}$. This is straightforward to see using the dark state $\ket{\mathcal{D}(t)}$ defined in Eq.\,(\ref{eq:dark}).
When the turn-off condition is modified such that the pump and Stokes pulses satisfy $\Omega_s(t)/\Omega_p(t) \rightarrow 1$, the superposition state $(\ket{1} -\ket{-1})/\sqrt{2}$ is created.
This is achieved without changing $\Omega_e(t)$ by simply using $\theta(t)/2$ (which is swept from $0$ to $\pi/4$ instead of $0$ to $\pi/2$) as the mixing angle. This results in new pulse shapes for $\Omega_p(t)$ and $\Omega_s(t)$ according to the following equations
\begin{align}\label{eq:HSTIRAP}
\begin{split}
\Omega_p(t) &= \Omega_{e}(t) \sin\left(\tfrac{1}{2}\theta(t)\right) \,,\\
\Omega_s(t) &= \Omega_{e}(t) \cos\left(\tfrac{1}{2}\theta(t)\right) \,.
\end{split}
\end{align}
We call this protocol half-STIRAP (H-STIRAP).
It is interesting to note that the non-adiabatic coupling $\dot{\theta}$ in the H-STIRAP protocol is exactly half of that in the non-adiabatic coupling in the STIRAP protocol~\cite{Bergmann_STIRAP_Review}.
When using H-STIRAP pulses to perform Ramsey interferometry, the second H-STIRAP pulse is time reversed (see Appendix~\ref{sec:Appendix:HSTIRAP}).



\subsection{Time dependence of Ramsey signal}

Figure~\ref{fig:H_STIRAP}(a) shows the DQ Ramsey interferometry pulse sequence, which is described as follows.
The $^{14}\mathrm{N}$ nuclear spins are polarized into the $\ket{+1}$ state  using a green laser pulse.
Next, the spin state is prepared in the superposition $\ket{\psi} = \left(\ket{+1} + e^{i\phi}\ket{-1}\right)/\sqrt{2}$ using two RF pulses:
a $\pi$-pulse on $f_1$ (duration $\pi/\Omega_p \approx \SI{18}{\micro s}$, frequency $\omega_p$),
which transfers population from $\ket{+1}$ to $\ket{0}$ and a two-tone pulse (duration $\pi/\Omega_e \approx \SI{13}{\micro s}$) with frequencies $\omega_p$ and $\omega_s$, which transfers the state from $\ket{0}$ to $\ket{\psi}$.
After a free precession time interval $\tau$, a second (identical) two-tone RF pulse is then used to convert the accumulated phase into a population difference, which is read out optically.
Figure~\ref{fig:H_STIRAP}(b) shows the adiabatic version of the DQ Ramsey interferometry pulse sequence, using half-STIRAP RF pulses whose construction is defined in Eq.\,(\ref{eq:HSTIRAP}). 
The first half-STRAP pulse adiabatically transfers the spin state from $\ket{+1}$ to $\ket{\psi}$, and a reverse half-STIRAP pulse is used to project the relative phase into a population difference.
For both pulse sequences (Fig.\,\ref{fig:H_STIRAP}(a) and Fig.\,\ref{fig:H_STIRAP}(b)), we employ a four-phase measurement (4-Ramsey)~\cite{Jarmola2021,Bar2020}, in which the phases of the final pulse ($\omega_p$ and $\omega_s$) are inverted ($+x \mapsto -x$), cycling through all four possible combinations (Fig.\,\ref{fig:H_STIRAP}(a) inset: $R_1$, $R_2$, $R_3$, $R_4$ and Fig.\,\ref{fig:H_STIRAP}(b) inset: $S_1$, $S_2$, $S_3$, $S_4$).

Figure~\ref{fig:H_STIRAP}c (and Fig.\,\ref{fig:H_STIRAP}d) shows optically detected Ramsey fringes from the $^{14}\mathrm{N}$ nuclear spin DQ transition, obtained by scanning $\tau$ using the pulse sequence shown in Fig.\,\ref{fig:H_STIRAP}a (and Fig.\,\ref{fig:H_STIRAP}b). 
This scanning of $\tau$ is performed in the lab frame, rather than in the rotating frame, resulting in oscillations that are detected at the transition frequency $f_1 - f_2$. In fact, scanning $\tau$ in the rotating frame would produce no oscillations, since we have no two-photon detuning ($\Delta_p - \Delta_s = 0$).
The top-left panels of both Fig.\,\ref{fig:H_STIRAP}c and Fig.\,\ref{fig:H_STIRAP}d show the Ramsey fringes for each of the four phase combinations for DQ 4-Ramsey ($R_1$, $R_2$, $R_3$, $R_4$) and STIRAP 4-Ramsey ($S_1$, $S_2$, $S_3$, $S_4$), respectively.
We find that when compared with individual DQ 4-Ramsey, the individual STIRAP 4-Ramsey measurements have a larger amplitude ($\sim 30\%$), reduced ``ripple'' ($\sim 1/20$), and an inverted phase.
The Fourier transforms of these individual signals (top-right panel of Fig.\,\ref{fig:H_STIRAP}c and Fig.\,\ref{fig:H_STIRAP}d) reveal that the ``ripple'' occurs at frequencies $f_1$ and $f_2$, corresponding to residual population in $\ket{0}$ due to pulse imperfections.
Combining the four individual Ramsey measurements,
($R = R_1 - R_2 + R_3 - R_4$) suppresses these residual signals, (bottom-left panel of Fig.\,\ref{fig:H_STIRAP}c and Fig.\,\ref{fig:H_STIRAP}d).
The suppression of the residual signals is also observed in the Fourier-transform plots (bottom-right of Fig.\,\ref{fig:H_STIRAP}c and Fig.\,\ref{fig:H_STIRAP}d).
The difference in amplitude (including the sign change) between DQ 4-Ramsey and STIRAP 4-Ramsey can be explained by the particular pair of states between which oscillations occur. In DQ 4-Ramsey, the amplitude is determined by the difference in brightness between $\ket{0}$, and $\left(\ket{+1} + e^{i\phi}\ket{-1}\right)/\sqrt{2}$, while in STIRAP 4-Ramsey, the amplitude is determined by the difference in brightness between $\ket{+1}$, and $\left(\ket{0} + e^{i\phi}\ket{-1}\right)/\sqrt{2}$.


\begin{figure}
\centering
    \includegraphics[width=1.0\textwidth]{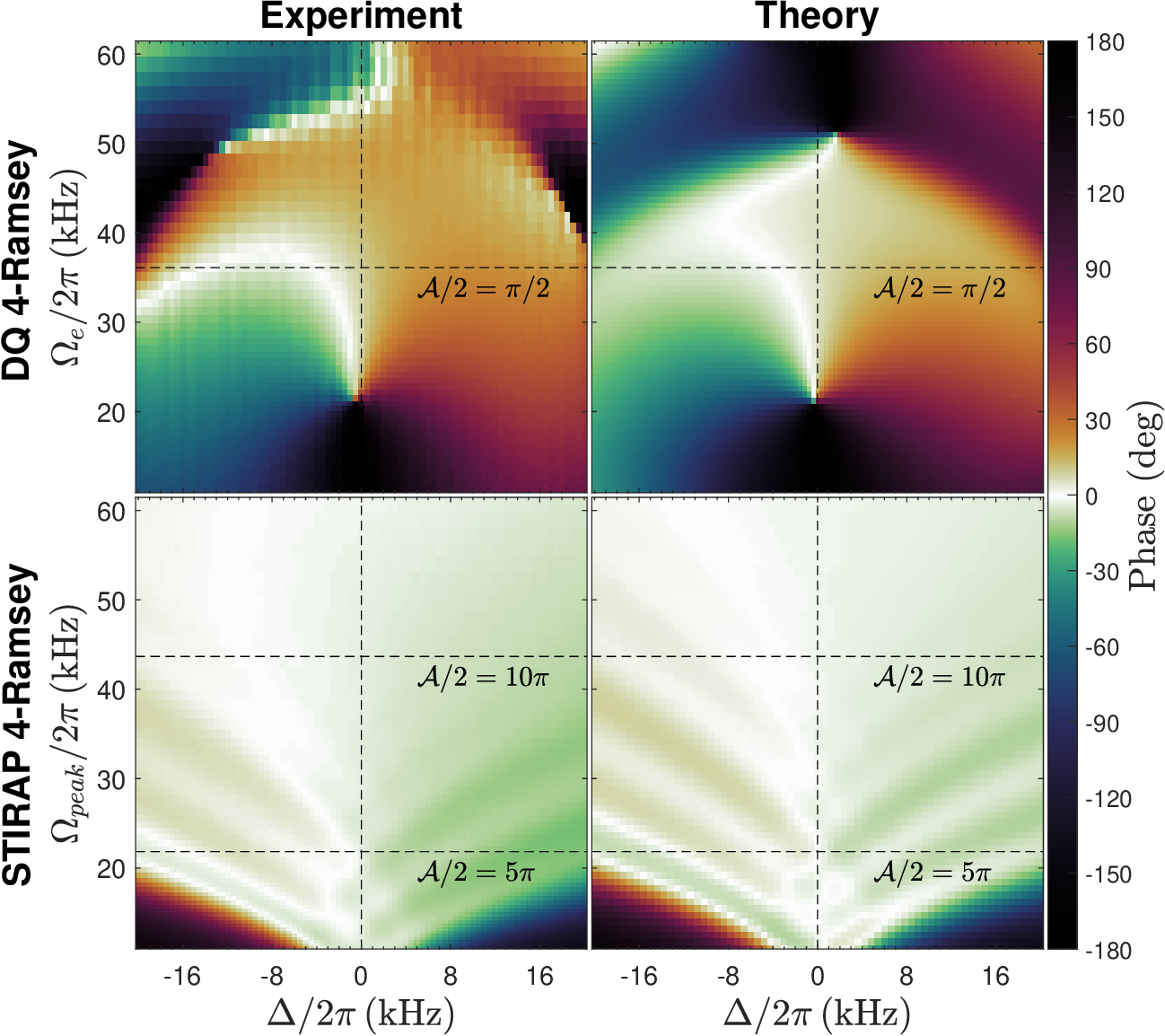}
    \caption{\label{fig:Robustness} \textbf{Robustness of DQ 4-Ramsey and STIRAP 4-Ramsey.}
    The accumulated phase ($\Phi$) of DQ 4-Ramsey and STIRAP 4-Ramsey interferometer techniques is plotted as a function of two parameters: detuning ($\Delta$) and effective Rabi frequency ($\Omega_e$).
    The left column shows experimentally measured results, and the right column shows theoretical predictions.
    For both techniques, the duration $t_p$ and separation $\tau$ of the pulses were fixed.
    For DQ 4-Ramsey (top row) $\tau = \SI{1.2}{ms}$, $t_p = \SI{13.9}{\micro s}$, and the horizontal dashed line corresponds to tuned pulse areas for all three pulses (Fig.\,\ref{fig:H_STIRAP}) $\mathcal{A}/2 = \pi/2$, $\Omega_e = 2\pi \times \SI{36.1}{kHz}$. 
    For STIRAP 4-Ramsey (bottom row) $\tau = \SI{0.8}{ms}$, $t_p = \SI{500}{\micro s}$, and the horizontal dashed lines indicate where the pulse area is $\mathcal{A}/2 = 5\pi$ and $10\pi$.
    }
\end{figure}

\subsection{\label{sec:Robustness}Robustness}

We characterize the robustness of the measurement of the DQ transition frequency ($f_1 - f_2$) against detuning ($\Delta$) and Rabi frequency ($\Omega_e$) for the two techniques shown in Fig.\,\ref{fig:H_STIRAP}: DQ 4-Ramsey and STIRAP 4-Ramsey. These two parameters are of particular interest for robustness because they are known to fluctuate when subjected to drifts in ambient temperature.

The accumulated phase $\Phi$ was obtained by varying the phase of the final Ramsey pulse. More precisely, $\Phi = \tan^{-1}\big(Q/I\big)$, where $I$ is the measurement obtained as previously described in Fig.\,\ref{fig:H_STIRAP}a and Fig.\,\ref{fig:H_STIRAP}b, and $Q$ is obtained in the same way, but shifting the DQ phase (relative phase between pump and Stokes) of the final RF pulse by 90 degrees
(see Appendix~\ref{sec:Appendix:Phase}).

Figure~\ref{fig:Robustness} shows 2D color plots of the accumulated phase $\Phi$ as a function of $\Delta$ and $\Omega_e$, obtained using both theory and experiment for both DQ 4-Ramsey and STIRAP 4-Ramsey.
It should be noted that the pulse durations, not pulse areas, are fixed for these plots. 
For DQ 4-Ramsey, the pulse durations were chosen to have optimal pulse areas (e.g. $\mathcal{A}/2 = \pi/2$) when $\Omega_e/2 \pi = \SI{36.1}{kHz}$, indicated by the horizontal dashed line. 
For STIRAP 4-Ramsey, the accumulated phase $\Phi$ is plotted against $\Omega_{peak}$, the peak value of $\Omega_e(t)$.

As expected, we found that the range of conditions over which the interferometer performs well is broader for STIRAP 4-Ramsey than for DQ 4-Ramsey.
It is interesting to note that the distribution of Rabi frequencies (due to RF gradients) actually improves the robustness of STIRAP 4-Ramsey, smoothing out the diagonal stripes, especially for large pulse areas.

We found that for DQ 4-Ramsey, the accumulated phase also depends greatly on the phase of the initial $f_1$ $\pi$-pulse. 
Because the $^{14}\mathrm{N}$ nuclear spins are initially polarized in the $\ket{+1}$ state, STIRAP 4-Ramsey does not require any initial pulse, and benefits in robustness as a result.
Note that in the electron-spin triplet of the NV center, the situation is reversed: polarization occurs in $\ket{m_S =0}$ and therefore STIRAP 4-Ramsey requires an initial $\pi$ pulse and DQ 4-Ramsey does not~\cite{Bohm2021_NV_STIRAP,Gong2024_NV_STIRAP}.

Under our experimental conditions
($\Omega_e/2 \pi = \SI{36.1}{kHz}$,
$\Delta = 0$),
we measure the dependence of the accumulated phase on detuning (first derivative) to be $\sim$~\SI{2.1}{deg \per kHz} for DQ 4-Ramsey.
This corresponds to phase drift of $\sim$~\SI{1.3}{\milli \radian \per K}, which follows from the temperature dependence of the $f_1$ and $f_2$ transitions (\SI{-35}{Hz/K},~\cite{Lourette2023}).
As a result, we expect this level of phase drift to ultimately limit the bias stability of rotation sensing with $^{14}\mathrm{N}$ nuclear spins.
STIRAP 4-Ramsey was measured to be more robust than DQ4R against changes in $\Delta$ under
the same
experimental conditions by a factor of 5,
which can be further improved by
choosing a value of pulse area for which this dependence vanishes.

\section{\label{sec:Conclusions}Conclusions and outlook}

In this work, we have experimentally demonstrated STIRAP in
the $^{14}\mathrm{N}$
nuclear hyperfine manifold in the ground state of NV centers in diamond driven by RF pulses. 
A new method of measuring nuclear-spin-state population dynamics (specific for NV centers in diamond) was developed and implemented.
We demonstrate a substantial suppression of the intermediate state population during STIRAP transfer between the $^{14}\mathrm{N}$ nuclear spin states. The experimental results are in good agreement with the developed theory, taking into account both gradients of the RF field across the measured volume as well as ac Zeeman shifts related to off-resonant components of the RF field excitation.

Utilizing the advantages of the STIRAP protocol, we developed and implemented an advanced Ramsey interferometric method based on the half-STIRAP pulse sequence. Effectively, we replace the traditionally used $\pi/2$-pulses in basic Ramsey schemes with half-STIRAP pulses. This modification allows for robust preparation
of the state superposition
and readout of the generated coherence.
We compared the performance of STIRAP Ramsey with the performance of the DQ Ramsey scheme, which is based on the effective $\pi/2$-pulse sequence, as in the standard Ramsey protocol.
Our findings indicate that STIRAP Ramsey offers enhanced tolerance to moderate changes in applied pulse parameters, resulting in improved robustness of Ramsey signal phase and amplitude (contrast).

The results show that STIRAP and its variations can effectively manipulate nuclear state populations and coherences, driving progress in quantum sensing applications like rotation sensing~\cite{Jarmola2021,Soshenko2021,AjoyPRA2012}, secondary frequency standards~\cite{FrequencyStandardBook}, quantum memory devices~\cite{Arunkumar2023_QuantumLogic}, and other quantum technologies. The technique's robustness against experimental imperfections will benefit future sensing and spectroscopy advances. Additionally, the $\pi/2$-pulse to half-STIRAP pulse conversion can be applied to external degrees of freedom in atom interferometry~\cite{Berman1997}, inheriting the robustness benefits. This STIRAP Ramsey scheme can be extended to other spin and atomic systems, providing improvements in robustness.

\section*{\label{sec:Ackno}Acknowledgments}
The authors are grateful to Alexander Pines, Malcolm Levitt, Dieter Suter, Victor M. Acosta, and Genko Genov for stimulating  discussions.
S.L., A.J. S.C., and J.C. acknowledge support from the DEVCOM Army Research Laboratory under Cooperative Agreement No. W911NF-24-2-0050, No. W911NF-18-2-0037, No. W911NF-24-2-0044, and No. W911NF-22-2-0097. S.A.M. acknowledges support from the Office of Naval Research under Awards No. N00014-20-1-2086 and N00014-22-1-2374, as well as from the Helmholtz Institute Mainz Visitor Program and the Alexander von Humboldt Foundation. D.B. acknowledges support from the DEVCOM Army Research Laboratory under Cooperative Agreements No. W911NF2120180, W911NF2320093.

\appendix

\section{\label{sec:Appendix:Experiment}Experimental setup details}

A 0.79–numerical aperture aspheric condenser lens (Thorlabs, ACL25416U-A) was used to illuminate a spot of diameter $\SI{\sim 50}{\micro m}$ on the diamond with $\SI{\sim 30}{mW}$ of 532 nm laser light (Coherent Verdi G5) and collect fluorescence.
Laser pulses were generated passing a continuous wave beam through an acousto-optic modulator.
The NV sensing volume is
$\sim 0.4 \,\si{nL}$
($\sim \qtyproduct{50 x 50 x 150}{\micro m}$),
defined by the area of the incident laser beam and the length of its path through the diamond.
The fluorescence was separated from the excitation light by a dichroic mirror, passed through a band-pass filter (650 to 800 nm), and detected by a free-space Si photodiode.

Radio-frequency (RF) pulses were generated by an RF synthesizer (Keysight 33512B) using arbitrary waveform generation and amplified using a broadband RF amplifier (Minicircuits LZY-22+). Microwave (MW) signals were generated by a MW synthesizer (Rohde \& Schwarz SMW200A), formed into pulses using an RF switch (Minicircuits ZASW-2-50DR+), and amplified using a broadband MW amplifier (Minicircuits ZHL-50W-63+).
RF and MW signals were combined using a diplexer (MarkiMicrowave DPX-0R5) and delivered using a 160-\si{\micro m}-diameter copper wire placed on the diamond surface next to the optical focus.

A TTL pulse card (SpinCore, PBESR-PRO-500) was used to generate and synchronize the pulse sequence. A data acquisition card (National Instruments, USB-6361) was used to digitize experimentally measured signals.

\section{\label{sec:Appendix:HSTIRAP}Half-STIRAP pulse construction}
\begin{figure}
\centering
    \includegraphics[width=1.0\textwidth]{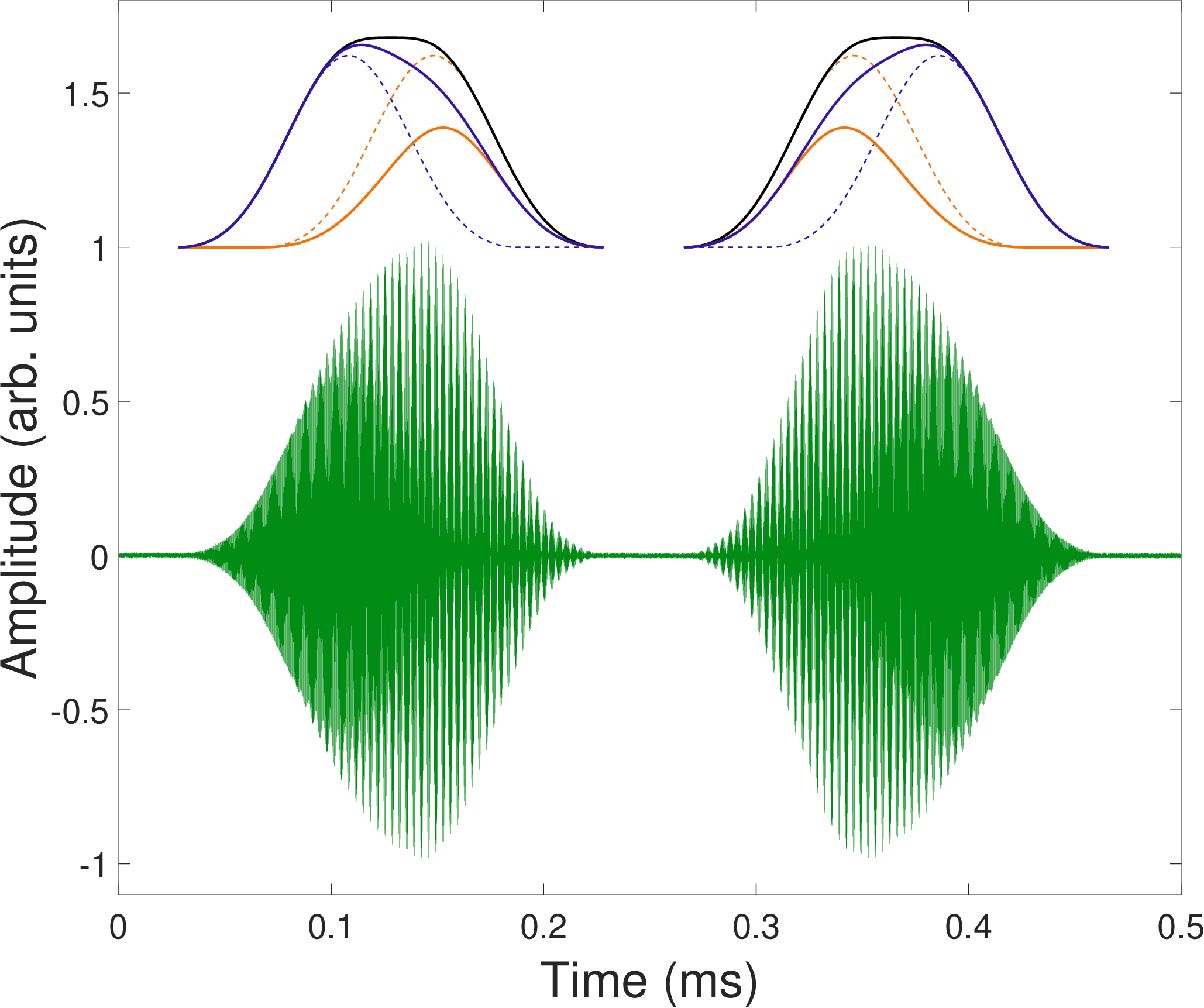}
    \caption{\label{fig:Waveform} \textbf{Waveforms of the half-STIRAP pulses used in STIRAP Ramsey.}
    The waveforms of the applied RF pulses for STIRAP Ramsey as used in the experiment, shown here with a reduced value of pulse separation $\tau$, and for $t_p = \SI{200}{\micro s}$.
    The waveforms of the half-STIRAP pulses are shown in green, captured using an oscilloscope.
    The visible oscillations are a result of the the pump and Stokes pulses beating against one another at their difference frequency.
    Plots of the pulse amplitudes ($\Omega_p(t)$ and $\Omega_s(t)$) of the pump (orange) and Stokes (blue) components are shown above.
    The orange and blue dashed lines show the pump and Stokes pulses for the original Blackman-shaped STIRAP pulses, respectively. Both sets of pulses share the same effective Rabi frequency $\Omega_e(t)$, shown in black.
    }
\end{figure}

The construction of the half-STIRAP pulses is described in Eq.\,\eqref{eq:HSTIRAP}.
Figure~\ref{fig:Waveform} shows the waveforms of the half-STIRAP pulses captured using an oscilloscope (green), in addition to plots of the Rabi frequencies of the pump (orange) and Stokes (blue) components.

\begin{figure}
\centering
    \includegraphics[width=1.0\textwidth]{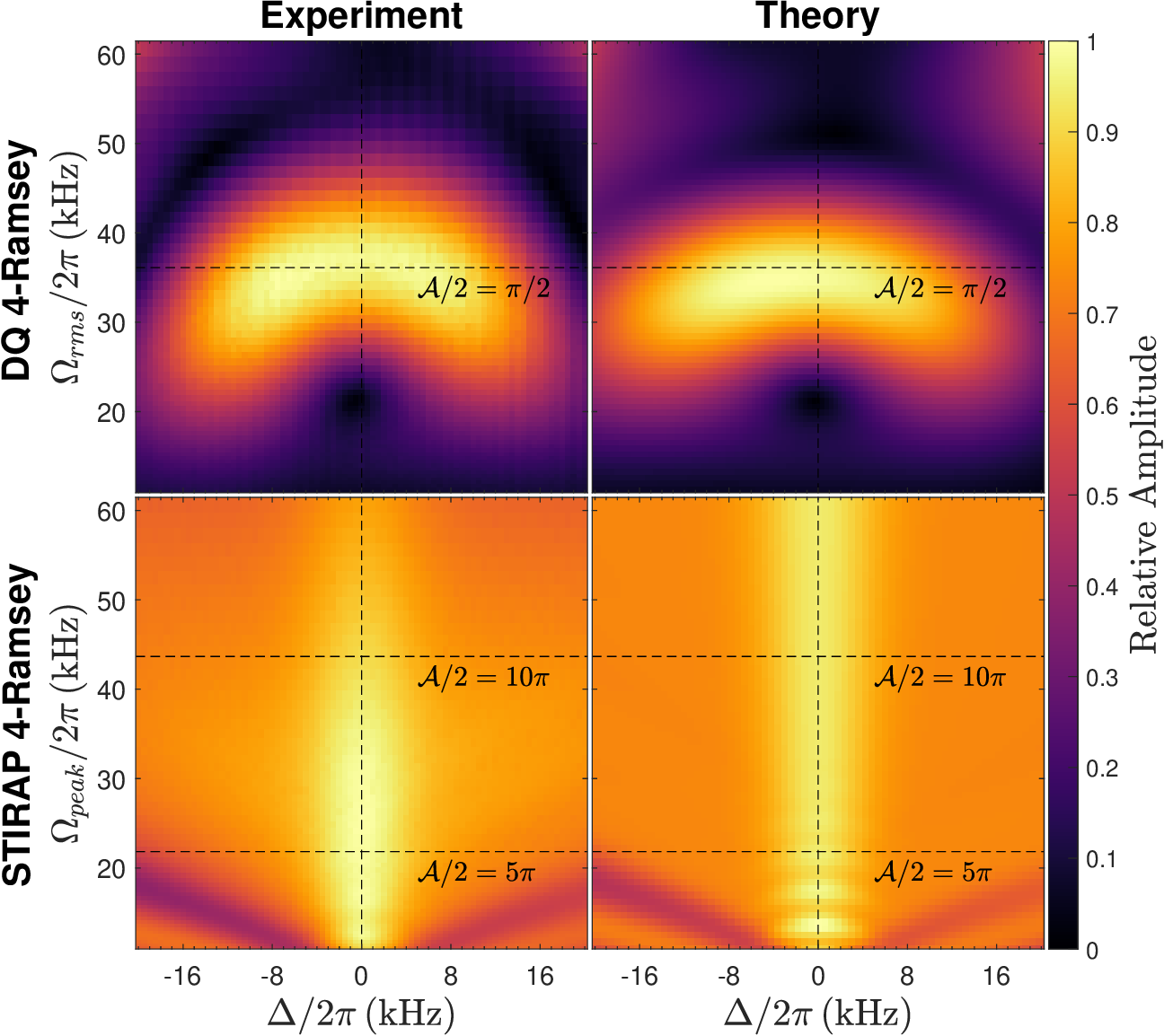}
    \caption{\label{fig:Robustness_Amplitude} \textbf{Robustness of the amplitude of DQ 4-Ramsey and STIRAP 4-Ramsey.}
    Amplitude plot of DQ 4-Ramsey and STIRAP 4-Ramsey interferometer techniques. While Fig.\,\ref{fig:Robustness} shows the accumulated phase $\Phi$, this plot shows the corresponding amplitude $r$ (defined in Eq.\,\eqref{eq:complex}) for the same data set.
    }
\end{figure}

\section{\label{sec:Appendix:Phase}Accumulated Phase measurements}

As described in Fig.\,\ref{fig:H_STIRAP}, the phase of the final Ramsey pulse is modulated to cancel out the effects of single-quantum coherence, both for DQ 4-Ramsey and STIRAP 4-Ramsey. To describe the process in greater detail, we define the phases of the first Ramsey pulse to be $\phi_{p1}$ for pump and $\phi_{s1}$ for Stokes, and we define the phases of the second Ramsey pulse to be $\phi_{p2}$ for pump and $\phi_{s2}$ for Stokes. In the non-rotating frame, pulses are then constructed according to
$\widetilde{\Omega}(t) = \mathrm{Real}\left[\Omega(t) e^{ i (\phi + \omega t)}\right]$,
where $\omega$ and $\phi$ are the respective frequency and phase of the RF pulse.

The phases of each pulse in DQ Ramsey are shown in the following table

\begin{center}
\begin{tblr}
{
    colspec={|c || c c c c | c c c c ||},
    colsep = {4pt}  
}
 \hline
 & \SetCell[c=4]{}I  &&&& \SetCell[c=4]{} Q \\[0.5ex] 
 \hline
  & $R_1$ & $R_2$ & $R_3$ & $R_4$
  & $R_5$ & $R_6$ & $R_7$ & $R_8$  \\ 
 \hline\hline
 $\phi_{p1}$
 & 0 & 0 & 0 & 0
 & 0 & 0 & 0 & 0  \\ 
 \hline
 $\phi_{s1}$
 & 0 & 0 & 0 & 0
 & 0 & 0 & 0 & 0 \\
 \hline
 $\phi_{p2}$
 & $0$
 & $\pi$
 & $\pi$
 & $0$
 & $0   + \frac{\pi}{4}$
 & $\pi + \frac{\pi}{4}$
 & $\pi + \frac{\pi}{4}$
 & $0   + \frac{\pi}{4}$
 \\
 \hline
 $\phi_{s2}$
 & $0$
 & $0$
 & $\pi$
 & $\pi$
 & $0   - \frac{\pi}{4}$
 & $0   - \frac{\pi}{4}$
 & $\pi -\frac{\pi}{4}$
 & $\pi -\frac{\pi}{4}$,
 \\ 
 \hline
 \end{tblr}
\end{center}

where $R_{1-4}$ are used to measure $I$, and $R_{5-8}$ are used to measure $Q$, according to the equations
\begin{align}
\begin{split}
    I &= R_1 - R_2 + R_3 - R_4 \, ,\\
    Q &= R_5 - R_6 + R_7 - R_8 \, .
\end{split}
\end{align}
The opposite $\pm\pi/4$ shifts in the second Ramsey pulse correspond to a $\pi/2$ phase shift in the effective DQ phase, $\phi_p - \phi_s$.

For STIRAP Ramsey, the procedure is identical except $S_{1-8}$ are used in place of $R_{1-8}$. For both DQ Ramsey and STIRAP Ramsey, the accumulated phase $\Phi$ (and amplitude $r$, shown in Fig.\,\ref{fig:Robustness_Amplitude})
is obtained by combining $I$ and $Q$ into a complex number, according to
\begin{equation} \label{eq:complex}
    z = I + iQ = re^{i\Phi} \, .
\end{equation}

\begin{figure}[t]
\centering
    \includegraphics[width=1.0\textwidth]{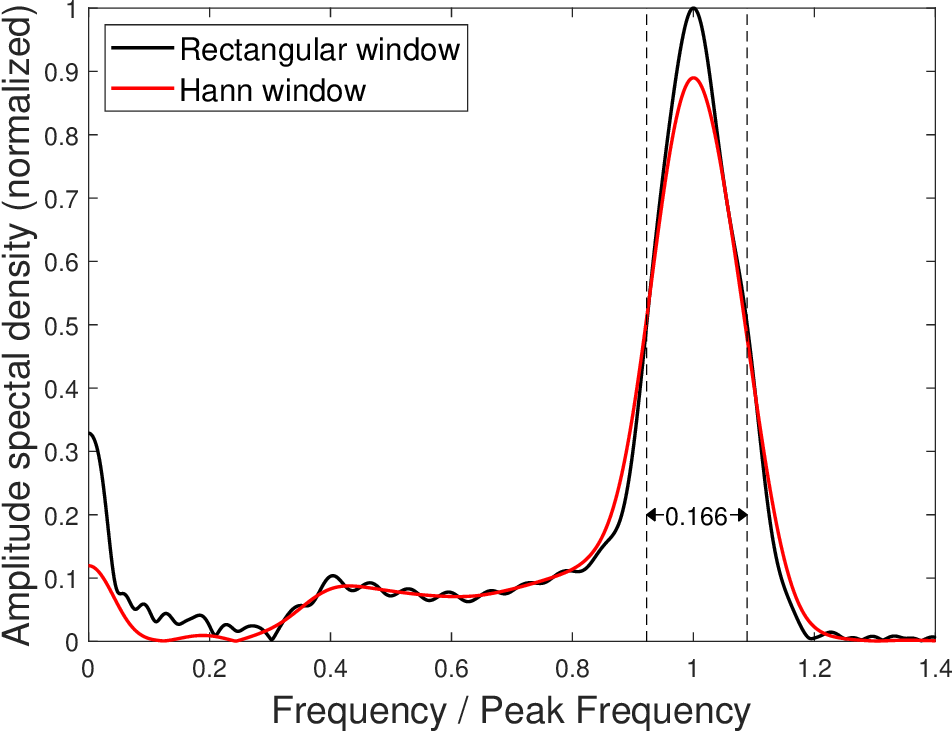}
    \caption{\label{fig:Distribution} \textbf{Experimentally measured distribution of Rabi frequencies.} The distribution of Rabi frequencies was obtained by scanning the duration of a rectangular RF pulse on the $f_1$ transition and performing Fourier analysis on the optically detected signal. The resulting amplitude spectrum is shown for both rectangular and Hann windows. The full width at half maximum (FWHM) was measured to be $0.166 \left<\Omega_{e}\right>$.
    }
\end{figure}

\section{\label{sec:Appendix:Simulations}Procedure for Numerical Simulations}

{\it Zeeman Shifts.} 
In the simulations, to account for the ac Zeeman shifts, we modify the Hamiltonian in Eq.\,(\ref{eq:Hamiltonian}) as
\begin{align}\label{AC_Hamiltonian}
\bar{H} (t)= \frac{\hbar}{2}
\left(\begin{array}{ccc}
 0 & \bar{\Omega}_p(t)  & 0 \\
\bar{\Omega}^*_p(t)  & -2\Delta & \bar{\Omega}_s(t) \\
0 & \bar{\Omega}^*_s(t) & 0 
\end{array}\right) \,,
\end{align}
where $\bar{\Omega}_p(t) = \Omega_p(t) + \Omega_s(t) e^{-i\eta(t)}$, $\bar{\Omega}_s(t) = \Omega_s(t) + \Omega_p(t) e^{-i\eta(t)}$, 
$\eta(t) = \left(\omega_p - \omega_s\right)t + \phi_p -\phi_s$
with $\phi_p$ and $\phi_s$ being the initial phases of the RF pulses. The modified Rabi frequencies take into account that both pump and Stokes fields interact on both transitions in the three-level system. Effectively, this leads to the modulation of the Rabi frequency envelopes with a modulation frequency equal to the difference $\omega_p - \omega_s$~\cite{MalQChem2012,MalSolSt2012}. A comparison between calculations with and without this correction reveals that the difference is small but noticeable. 

{\it State propagation}. We obtain the numerical results in Fig.\,\ref{fig:FSTIRAP} by solving the von Neumann equation. The initial density matrix is a mixed state matching the experimentally measured probabilities at $t=0$. In Fig.\,\ref{fig:Robustness} simulation, we assume that the initial state is $\ket{1}$ and solve the Schr\"odinger equation.
In both cases,
the condition of two-photon resonance is assumed, $\delta=\omega_p - \omega_s - \omega_1 + \omega_2=0$. The experimental uncertainty associated with the two-photon detuning is approximately 5 Hz.
Theoretical modeling and simulations were performed using Julia's QuantumControl.jl package~\cite{GoerzQ2022}, which uses Chebyshev's polynomials to calculate the time evolution.


{\it Rabi frequency gradient.} We average an ensemble of simulations with different Rabi frequencies to account for RF power gradients across the sensing volume. We take those Rabi frequencies from a Gaussian distribution centered in the reported Rabi frequencies in the main text. The width of the Gaussian distribution, the standard deviation, was verified experimentally and the results are shown in Fig\,\ref{fig:Distribution}. The average population of the state $\ket{i}$ at time $t$ is given by
\begin{equation}
    \langle p_i(t)\rangle =  \frac{1}{\sigma \sqrt{2\pi}} \int\limits_{-\infty}^\infty p_i(t, \Omega_e') \exp\left[-\frac{(\Omega_e'-\langle\Omega_e\rangle)^2}{2\sigma^2}\right]d\Omega_e' \, .
\end{equation}
where $\langle\Omega_e\rangle$ is the average peak amplitude of the effective Rabi frequency, $\sigma$ the standard deviation, and $p_i(t, \Omega_e')$ the population of the state $\ket{i}$ at time $t$ when simulating using a peak amplitude of the effective Rabi frequency $\Omega_e'$.

{\it Ramsey signal}. The spin-state wavefunction for the DQ Ramsey scheme is given by
\begin{equation}
    \ket{\psi_{f}} = U^{(2)}_{\pi/2} P_{\tau}U^{(1)}_{\pi/2} P U_{\pi} \ket{\psi_{0}}
\end{equation}
where $\ket{\psi_0}=\ket{0}$ is the initial state, $U_{\pi}$, $U^{(1)}_{\pi/2}$, $U^{(2)}_{\pi/2}$ are the evolution operators with the Hamiltonian in Eq.\,(\ref{AC_Hamiltonian}), the sub-indexes indicate the pulse area. 
The transformation matrices $P$ and $P_{\tau}$ take into account the phases between pulses and the free-evolution phase accumulated  by the states. 

For the H-STIRAP  Ramsey scheme, $\ket{\psi_0}=\ket{1}$, $U_{\pi} = P = \hat{I}$, and the  evolution operators $U^{(1),(2)}_{\pi/2}$  are calculated using pump and Stokes Rabi-frequency envelopes described in the main text, Eq.\,(\ref{eq:HSTIRAP}).  

The experimentally measured optical readout is the 
total fluorescence signal from all three nuclear spin states. To account for the 
relative brightness of the $^{14}\mathrm{N}$ nuclear states, we calculate the Ramsey signal as 
\begin{equation}
R = |\langle 1|\psi_f\rangle|^2 + 0.9785 |\langle 0|\psi_f\rangle|^2 + 0.9861 |\langle -1|\psi_f\rangle|^2 \,,
\end{equation}
where 1, 0.9785, and 0.9861 are the relative brightness coefficients
for the $\ket{1}$, $\ket{0}$, $\ket{-1}$ states respectively~\cite{Jarmola2020Robust}.





\bibliography{diamond}

\end{document}